\documentclass[12pt]{iopart}
\usepackage{amssymb}

\usepackage{graphicx}
\usepackage{color}

\begin{document}

\DeclareGraphicsExtensions{.pdf,.png,.gif,.jpg,.eps}
\title{Probing Interaction-Induced Ferromagnetism in Optical Superlattices}
\author{J. von Stecher,$^1$ E. Demler,$^{2,3}$ M. D. Lukin,$^{2,3}$ and A. M. Rey$^{1}$}
\address{$^1$JILA, University of Colorado and National Institute of Standards and Technology,
Boulder, Colorado 80309-0440,}
\address{$^2$Physics Department, Harvard University, Cambridge, Massachusetts, 20138,}
\address{$^3$Institute for Theoretical Atomic, Molecular and Optical Physics, Cambridge, Massachusetts 02138, USA
}

\begin{abstract}
We propose a  method for controllable  preparation  and detection of  interaction-induced ferromagnetism in ultracold fermionic atoms loaded in optical superlattices. First, we  discuss how to probe and control Nagaoka ferromagnetism in an array of isolated plaquettes (four lattice sites arranged in a square). Next, we allow  for  weak interplaquette tunneling. Since ferromagnetism is unstable in the presence of weak interplaquette couplings,  we propose to mediate long-range ferromagnetic correlations  via double-exchange processes  by exciting  atoms to the first vibrational band. We calculate   the phase diagram of the  two-band plaquette array  and discuss conditions for the stability and robustness of the ferromagnetic phases in this system. Experimental implementations   of the proposed schemes  are  discussed.

\end{abstract}

\maketitle

\section{Introduction}

The origin of ferromagnetism in itinerant electron systems remains an important open problem in condensed matter physics.
Mean field approaches, such as the  Hartree-Fock approximation~\cite{PhysRev.142.350}  and the Stoner criterion~\cite{stoner1938cef} for ferromagnetic instabilities, are extremely unreliable since they overestimate the stability of the magnetic-ordered phases~\cite{fazekas1990gsp}. The only rigorous example of ferromagnetism in the generic Hubbard model~\cite{Wu}, predicted by Nagaoka in 1965~\cite{Nagaoka1966}, was proven for a system with one fewer electron than half filling (i.e., one hole) in the limit of infinite interactions. Even though such a ferromagnetic state  is an iconic  example of a strongly correlated many-body state, it is highly unstable and  counter examples indicating the absence of ferromagnetism with two or more holes have been found~\cite{takahashi1982chm}.

The experimental observation of Nagaoka ferromagnetism (NF) is a challenging task, as it requires a system with a finite and controllable number of holes.
Even though there have been recent attempts to explore Nagaoka ferromagnetism using arrays of quantum dots~\cite{Nielsen2007},
the exponential sensitivity of the tunneling rates to the interdot distance and the random magnetic field fluctuations  induced by the nuclear spin background have  prevented its direct experimental observation.

 Here, we propose to use cold fermionic atoms in optical superlattices for the controllable observation of interaction-induced ferromagnetism.
First, we show how to probe the onset of NF in an array of  isolated plaquettes (four lattice sites in a square geometry).
Next, we discuss how  to engineer  long-range ferromagnetic correlations, starting from the isolated plaquette arrays.
Since weak coupling of the plaquettes destroys  ferromagnetism~\cite{Yao2007}, we instead propose to use additional atoms loaded in excited bands. The underlying idea is to use   Hund's rule couplings~\cite{hewson1993kph}  to favor local ferromagnetic alignment  among  the atoms in different bands and a double-exchange mechanism (tunneling-induced   alignment of the spins)~\cite{Zener} to stabilize ferromagnetic correlations between adjacent plaquettes. Exact numerical calculations for an array of weakly coupled plaquettes confirm the existence of stable ferromagnetic order in this two-band setup.
 Finally, we discuss methods for experimental preparation and detection of the ferromagnetic correlations.

\section{Probing Nagaoka ferromagnetism in superlattices.}

The low-energy physics of fermionic atoms loaded in the lowest vibrational band of  an optical lattice is well described by the Hubbard Hamiltonian:
\begin{equation}
\hat{H}= -  \sum _{\langle r,r'\rangle,\sigma} J_{r,r'} \hat{c}_{ r\sigma }^{\dagger}\hat{c}^{}_{r'\sigma}+U \sum_{r}
 \hat{n}_{\uparrow r}\hat{n}_{\downarrow,r}, \label{plaq}
\end{equation} where  $J_{r,r'}=J$ is the tunneling energy, and $U$ is the onsite interaction energy.
In Eq.~(\ref{plaq}), $\hat{c}^{}_{r\sigma}$ are fermionic annihilation operators, $\hat{n}_{r\sigma }=\hat{c}^{\dagger}_{r\sigma } \hat{c}^{}_{r\sigma }$ are number operators, $r=1, \dots L$ labels the lattice sites, and $\langle r,r'\rangle$ in the summation indicates that the sum is restricted to nearest neighbors.

Before starting let us  explicitly state  Nagaoka theorem \cite{Nagaoka1966}:
{\it ``Let the tunneling matrix element between lattice sites $r$ and $r'$ be negative,  $J_{rr'}<0$,   for any $r\neq r'$ and $U=\infty$ and let the number of fermions be $N=L-1$, with $L$ the total number of  sites. If the lattice satisfies certain connectivity condition, then the ground state has total spin $S=N/2$ and it is unique, apart  from a trivial $(N+1)$-fold spin degeneracy''}.

The notion of ``connectivity'' requires  that  each site in the lattice is contained in a loop (of non-vanishing $J_{rr'}$)  and
furthermore that the loops should pass through no more than four sites \cite{Tasaki1998}. The requirement $J_{rr'}<0$ is just the opposite of that assumed in most discussions of the Hubbard model.  However, in  bipartite lattices, there is always a canonical transformation connecting $J_{rr'}<0$ and $J_{rr'}>0$.

  From the conditions of the theorem  it is clear that the minimal geometry to observe Nagaoka ferromagnetism is a triangle, however a triangle is a trivial example since in this case either the ground state is always a singlet ( case  $J_{r,r'}=J>0$ ) or it is always a triplet {($J<0$) \footnote{Since a triangular lattice is not a bipartite lattice, Nagaoka theorem only holds for $J<0$.}. The first non trivial example of Nagaoka crossing takes place in a plaquette loaded with three fermions(Fig.~1(a)).

\begin{figure}[h]
\begin{tabular}{ccc}
\quad\quad\quad\quad\quad\quad\quad&
\includegraphics[scale=0.27,angle=0]{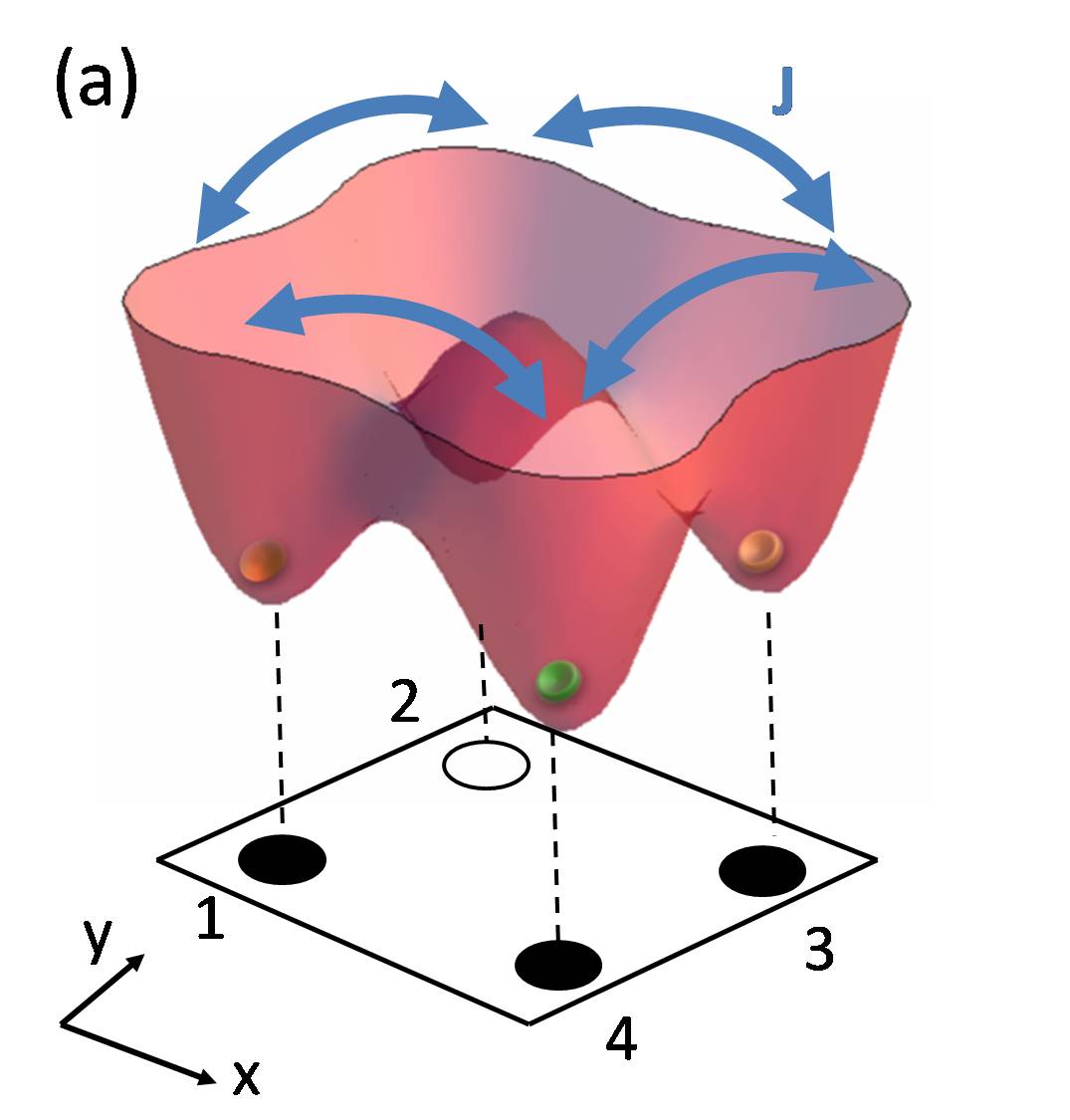} &
\includegraphics[scale=0.65,angle=0]{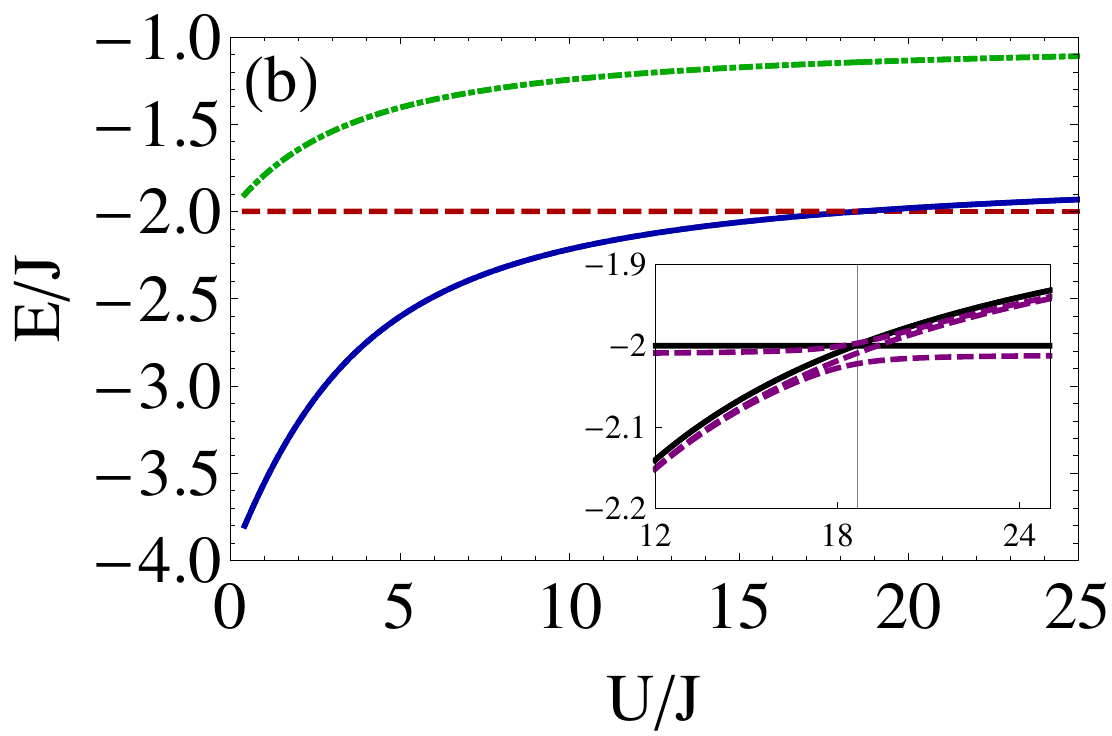}
\end{tabular}
\caption{ (a) Schematic representation of a plaquette. (b) Low energy spectrum of three fermions.
Solid lines and dash-dotted are $S=1/2$ while dashed curve corresponds to $S=3/2$. Inset: Zoom in of the spectrum at the Nagaoka crossing. Solid lines are the spectrum for zero gradient field while dashed lines are the spectrum at finite gradient field.}\label{fig1}
\end{figure}

The energy levels of a plaquette loaded with three fermions can be classified according to the total spin $S$ and  the symmetries of the wave function.
 It is known (e.g., Ref.~\cite{Yao2007}) that for
$U<U_t\approx 18.6 J$ the ground state is a degenerate doublet   $S=1/2$ state  with  $\tau=p_x \pm i p_y$  symmetry (the wave function changes phase by $\pm \pi/2$ upon $\pi/2$  rotation).  For  $U>U_t\approx 18.6\, J $, the ground state becomes a ferromagmetic $S=3/2$ state,  in agreement with the Nagaoka theorem (Fig.~\ref{fig1}(b)).
 We denote these eigenstates as $|S=1/2,S_z,\tau=\pm\rangle$ and $|S=3/2,S_z\rangle$ with $S_z=-S,\dots S$ and recall that the energies are independent of the $S_z$ value.
The onset of Nagaoka ferromagnetism can  be understood as competition between  the kinetic energy and superexchange interactions.
In the $U\to \infty$ limit, double occupancies are energetically suppressed,  and  the low-energy states are singly occupied with an energy spectrum given by  $E=\pm 2J, \pm\sqrt{3}J,\pm J, 0$. The relevant low-lying eigenstates are the ones with $E^{S=3/2}=-2J$ and $E^{S=1/2}=-\sqrt{3}J$. As $U$ become finite, while  the fully polarized states remain  eigenstates for any $U$ and their  energy is unaffected by interactions, the $ |S=1/2,S_z,\pm\rangle$  states acquire some admixture of double occupancies, which tend to lower their energy. The energy shift in the $S=1/2$ states  can be calculated by using second order perturbation theory, yielding $E^{S=1/2}= -\sqrt{3}J -\frac{5 J^2}{U} $. The Nagaoka crossing occurs at the $U_t/J$ value  when the two energies become equal, $U_t=5/(2-\sqrt{3}) J\sim 18.66 J$, in very good agreement with the exact diagonalization.

 An array of plaquettes can be created by superimposing two orthogonal optical superlattices formed by two independent sinusoidal potentials that differ in periodicity by a factor of two, i.e.,  $V(x)=V_s/2\cos(4 \pi x/\lambda_s)-V_l/8\cos(2 \pi x/\lambda_s)$, where $V_l$ is the long lattice depth,
 $V_s$ is the short lattice depth,
 and $\lambda_s$ is the short lattice wavelength. By controlling the lattice intensities, it is possible to tune the intra- and interplaquette tunneling and, in particular, to make the plaquettes independent. Here the axial optical lattice  is assumed to be deep enough to freeze 
any axial dynamics.
To load the plaquettes with three atoms, one can start by preparing a Mott insulator with filling factor three in a 3D lattice   and then slowly split the wells along x and y. Since only two fermions with opposite spin can occupy the lowest vibrational level, loading three fermions per site requires populating the first excited vibrational state  before splitting the wells. This loading procedure creates plaquettes with $S=1/2$.

Since $S$ and $S_z$ are  conserved quantum numbers in  clean cold atom set-ups,  to probe the Nagaoka transition we require the presence of a weak  magnetic-field gradient. We choose for this case a field pointing along $z$  with a constant gradient along the $x$ direction ${\bf{B}}(x)=\frac{\delta E_B}{\mu_B g}  \frac{2 x}{\lambda_s} {\bf{\hat{z}}}$, where $\mu_B$ is the Bohr magneton $g$ is the gyromagnetic factor.
The magnetic-field gradient couples the $ |3/2\rangle$ state with some linear combination of $|1/2,1/2,\pm\rangle$ which we denote as  $|1/2, 1/2,1\rangle$, through a Hamiltonian matrix element $H_{3/2,1/2}=-2/3 (1 + \sqrt{3}) \delta E_B$ and leaves another linear combination of $|1/2,1/2,\pm\rangle$, which we denote as $|1/2,1/2,2\rangle$, uncoupled. It consequently
transforms the crossing at $U_t$ into an avoided crossing [see inset of Fig.~\ref{fig1}~(b)] which   can be used to adiabatically transform the $|1/2,1/2,1\rangle$, ground state for $U<U_t$, into $ |3/2\rangle$ as $U$ is slowly increased.
$U$  can be tuned by a Feshbach resonance in the presence of an  external  magnetic field.

The variation of   $\langle S\rangle$ in the ground state below and above the Nagaoka point  can be inferred from a band mapping analysis \cite{Greiner2001} after  collapsing the plaquettes into  single wells:  The states with $S=3/2$ and $S=1/2$  will occupy  three and two bands respectively when a plaquette is transformed into a single site. Since the state $|1/2,1/2,2\rangle$ remains decoupled, one can
improve the band mapping signal by adiabatically  turning  on the magnetic field gradient at the same time the plaquettes are prepared. In this way the initial plaquette configuration will be  almost in a pure  $|1/2,1/2,1\rangle$ state and we will be able to fully transform the state into the $ |3/2\rangle$ at $U>U_t$.

The energy difference between the $|3/2\rangle$ and $|1/2,1/2,\pm\rangle$ states can also  be probed dynamically. After preparing    the state $|\psi(0)\rangle=\cos\alpha |3/2,1/2\rangle+\sin\alpha|1/2,1/2,1\rangle$ one can suddenly  turn off  the magnetic-field gradient. By  measuring the Neel
order parameter or spin imbalance along the $x$ direction [$N_S(t)=1/2(\sum_{r=1,2} n_{\uparrow r}-n_{\downarrow r}- \sum_{r=3,4}n_{\uparrow r}-n_{\downarrow r})$] as a function of time, one can track the Nagaoka point by the oscillation period of
$\langle N_S(t)\rangle\propto \cos[(E^{S=3/2}-E^{S=1/2})t/\hbar] $. As $U/J$ approaches  $U_t/J$, the period will become very long, indicating that the character of ground state has changed. This simple treatment ignores the admixture of  states with double occupied sites  in the $|1/2,1/2,1\rangle$ state. When included, the excitations introduce fast but small oscillations of frequency $J$. Comparisons between the exact and analytic solutions are shown in Fig.~\ref{fig3}.
The spin imbalance $N_S(t)$  can be experimentally probed by  first
splitting the plaquettes into two double wells and then following the same experimental methods used for measuring superexchange interactions~\cite{Trotzky2008}  that rely on band-mapping techniques and a Stern-Gerlach filtering.

\begin{figure}[h]
\begin{center}
\leavevmode {\includegraphics[scale=0.6]{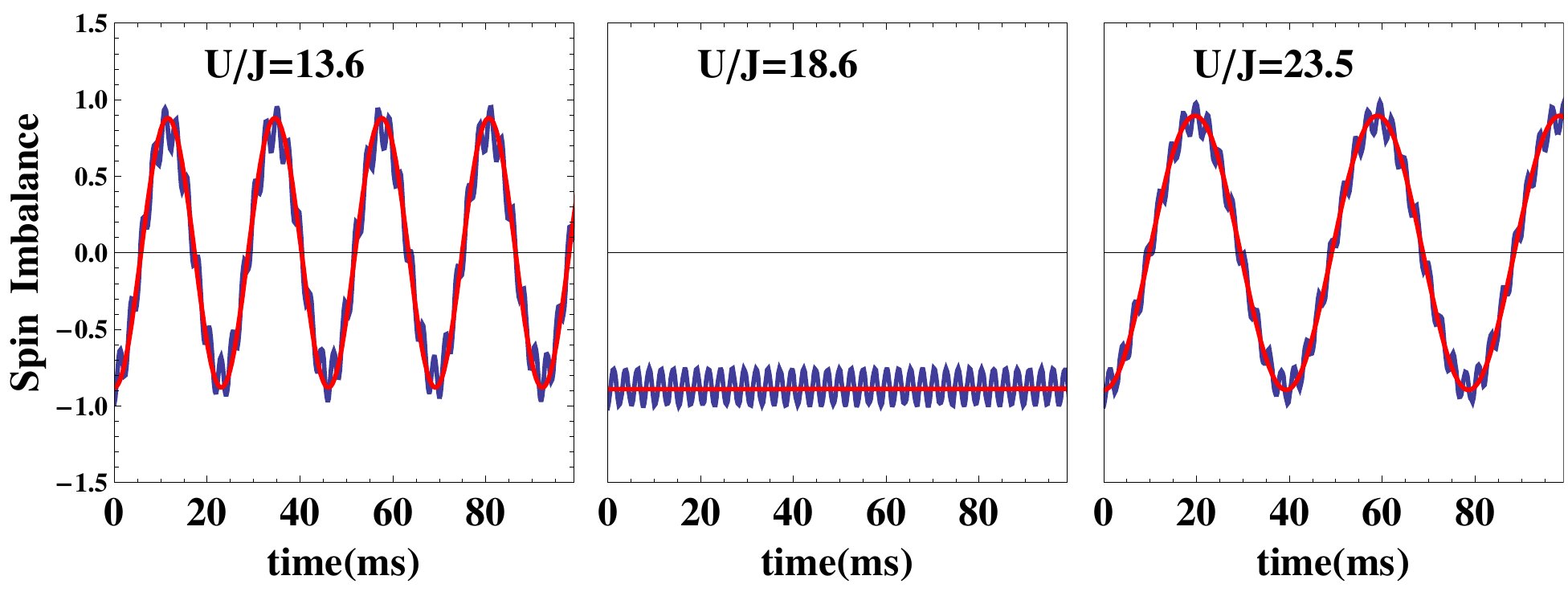}}
\end{center}
\caption{Normalized spin population imbalance. At the Nagaoka crossing, the envelope frequency becomes very long, indicating zero-energy splitting between the $|3/2\rangle$ and $|1/2\rangle$ levels.} \label{fig3}
\end{figure}

 The constant magnetic field needed for tuning a Feshbach resonance does not affect the dynamics since the relative energy spacing of the various levels within a plaquette is insensitive to such magnetic fields.
The big advantage of this probing method is that it  does not  require fixing the same magnetization for the various plaquettes. Consequently, we can relax the temperature constraint for preparing the Mott insulator used for the initial loading. The insensitivity of this probing method to the initial magnetization can be understood by the fact that the dynamic taking place in a plaquette initially loaded with $S_z=-1/2$ is identical to that described for the $S_z=1/2$ case. Furthermore, preparation of  plaquettes with $S_z=\pm3/2$ is energetically suppressed due to the large vibrational energy spacing. It should be stressed that finite temperature effects only determines the fidelity of the initial preparation of a Mott insulator. Once the Mott insulator is achieved, the observation of ferromagnetism depends on the efficiency of implementing the adiabatic manipulation discussed above.

\section{Engineering long-range ferromagnetic correlations. }

We now study the more general case in which one allows a weak interplaquette tunneling, $J'$, by lowering the  long lattice depth  along  both the $x$ and $y$ directions (or along only $x$). This procedure generates  a 2D (1D) array of plaquettes. In  the Nagaoka regime ($U/J>18.6$) to zero order in $J'$, the many-body  ground state has a degeneracy of $4^N$ ($N$ is the number of plaquettes) and is spanned by states of the form $|\Phi\rangle_{S_{z1},\dots S_{zN}}=\prod_i|S=3/2,S_{zi}\rangle$. A finite $J'$ breaks the degeneracy between the states, but as long as $J'\ll J$, the occupation of
states with $S_i<3/2$ is energetically
suppressed. These states can only be populated
 ``virtually,'' leading to an effective Heisenberg
interaction between the various effective $S=3/2$  states at each plaquette~\cite{Yao2007}, i.e.,
\begin{equation}
H_{eff}=G \sum_{\langle i,j\rangle}\vec{\hat{S}}_i\cdot \vec{\hat{S}}_j.\label{hei}
\end{equation} Here, $\vec{\hat{S}}_i=(\hat{S}_{xi},\hat{S}_{yi},\hat{S}_{zi})$ are spin $3/2$ operators acting on the pseudospin states $|S=3/2,S_{zi}\rangle$, and we have set $\hbar=1$. The interaction coefficient can be written as $G=g J'^2/J$, where
$g>0$ is an antiferromagnetic-coupling constant that slowly varies as a function of $J/U$.
  Equation~(\ref{hei}) explicitly shows the fragility of Nagaoka ferromagnetism, since a weak coupling among the plaquettes leads to a many-body ground state with antiferromagnetic correlations.

To overcome this limitation, we consider a different initial configuration. Starting with four atoms per plaquette in the lowest orbital, we excite one of the atoms to a nondegenerate vibrational level [see  Fig.~\ref{2pla2band}].
This system is described by a two-band Hubbard Hamiltonian of the form
 \begin{eqnarray}
\hat{H}&=& -  \sum _{\langle r,r'\rangle,\sigma,n} J_n \hat{c}_{ r n\sigma }^{\dagger}\hat{c}^{}_{r'n\sigma}+ \sum_{r n}
 U_{n,n}\hat{n}_{ r n \uparrow}\hat{n}_{r n\downarrow}+V\sum_{r }
 \hat{n}_{ r 1 }\hat{n}_{r 2 }\nonumber\\
 & &-J_{ex}\sum_{r\sigma \sigma'}\hat{c}_{ 1 r \sigma }^{\dagger}\hat{c}^{}_{1 r \sigma'}\hat{c}_{ 2 r \sigma' }^{\dagger}\hat{c}^{}_{2r\sigma}\label{H2b}
\end{eqnarray}
which is characterized by on-site interactions between particles in the ground ($U_{11}\equiv U$) and excited ($U_{22}\equiv U_e)$ bands, tunneling in the lower  ($J_{1}\equiv J$) and upper ($J_{2}\equiv J_e$) bands, and direct ($V$) and exchange ($J_{ex}$) interactions between the two bands. In the present implementation, $V=J_{ex}$. In Eq.~(\ref{H2b}), we have neglected terms that transfer atoms between bands, since they are energetically suppressed. The energy splitting between them has been omitted in our rotating frame.

\begin{figure}[h]
\begin{center}
\includegraphics[scale=0.7,angle=0]{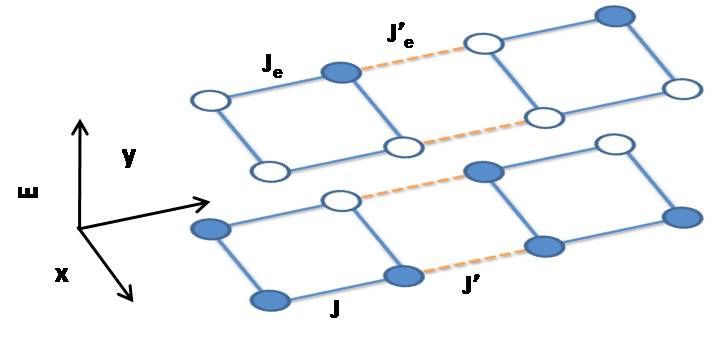}
\caption{  Schematic representation of two coupled double-band plaquettes. Solid circles represent occupied orbitals.}
\label{2pla2band}
\end{center}
\end{figure}

A single plaquette with three atoms in the lowest band and the fourth in the second band
exhibits a
crossing between an S=1 and an S=2 state (ferromagnetic state)
at a  value of $\tilde{U}$ that is smaller than $U_t$.
Figure~\ref{2pla}(a) shows the low energy behavior of this system. For the parameters used in Fig.~\ref{2pla}(a), the crossing occurs at $\tilde{U}\approx 6\, J$.
In general, $\tilde{U}$ depends on $J$, $J_e$, $U$, and $J_{ex}$. Figure~\ref{2pla}(b) analyzes the existence of a ferromagnetic ground state as a function of $U/J$ and $\alpha_J =J_e/J$. For this analysis, we assume that the interaction terms are proportional, i.e., $J_{ex}=\alpha_{J_{ex}} U$. Different shaded regions display the ferromagnetic regions for different $\alpha_{J_{ex}}$, and the dashed curves  their corresponding critical values  $\tilde{U}$. As both $\alpha_J$ and  $\alpha_{ex}$ increase, the $\tilde{U}$ values decrease, extending the ferromagnetic region. This behavior is consistent with our physical picture that both double exchange processes and the Hund's rule coupling stabilize ferromagnetism.

\begin{figure}[h]
\begin{center}
\begin{tabular}{cc}
\includegraphics[scale=0.46,angle=0]{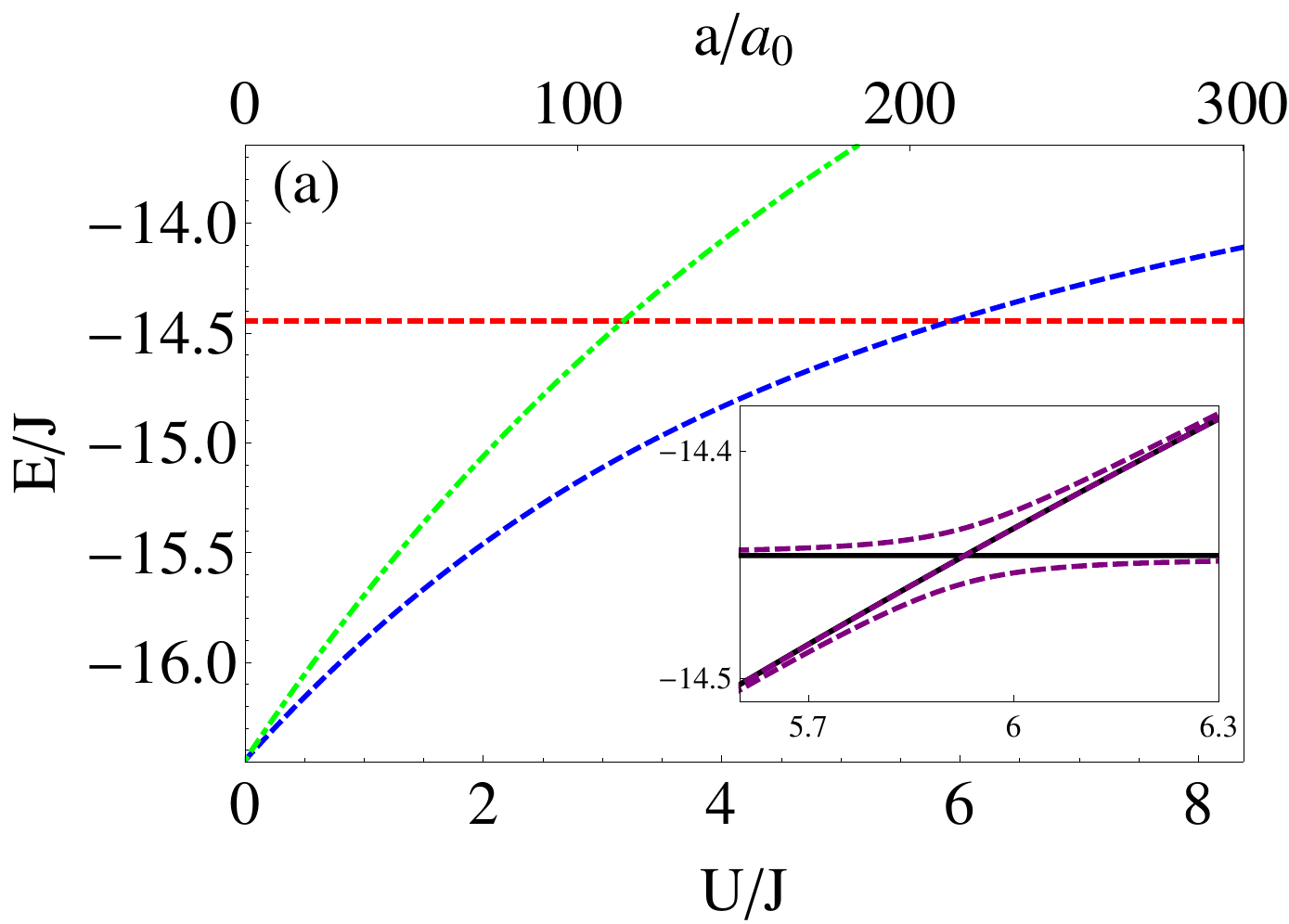} &
\includegraphics[scale=0.63,angle=0]{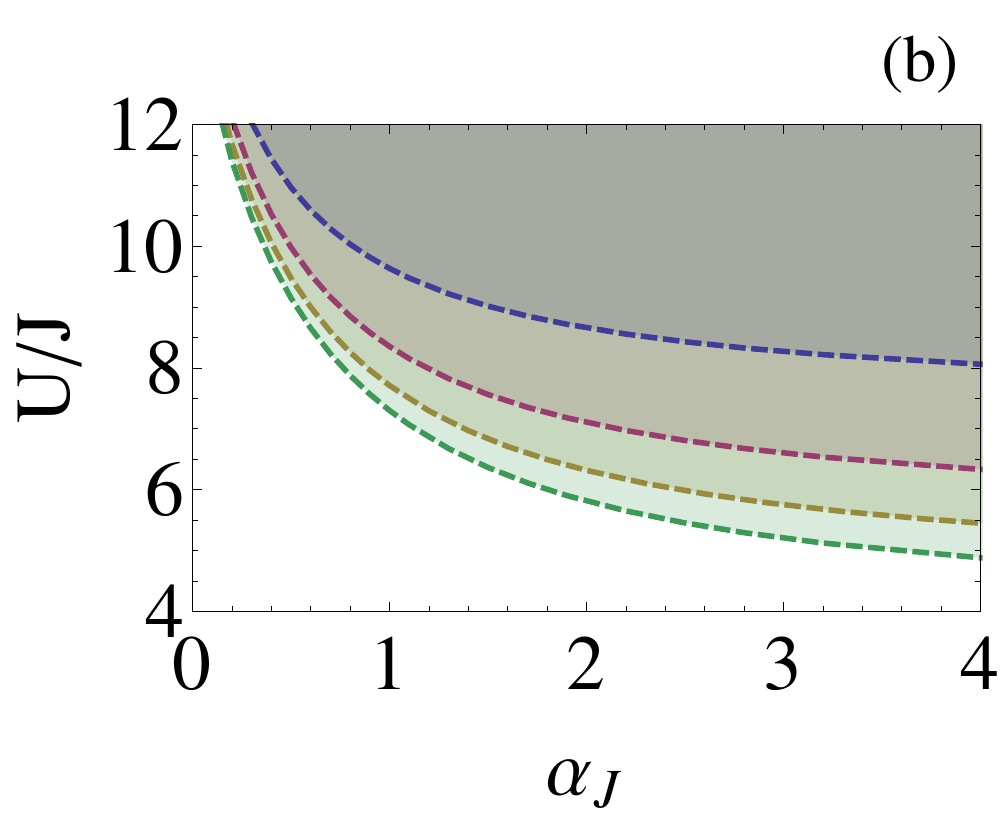}
\end{tabular}
\caption{ (a) Energies of a plaquette as a function of $U/J$ and the scattering length in Bohr radii $a_0$. The parameters that characterize the Hamiltonian [Eq.~(\ref{H2b})] are obtained for a superlattice constructed with a short-wavelength laser of $\lambda_s=765$ nm that characterizes the short-lattice recoil energy $E_r=h^2/(2m\lambda_s^2)$. The energies of the figure corresponds to $V_l=20\, E_r$ and  $V_s=7.5\, E_r$. Inset: Zoom in of the spectrum at the ferromagnetic crossing. Solid lines are the spectrum for zero gradient field while dashed lines are the spectrum at finite gradient field. (b) Ferromagnetism in a two-band plaquette. Shaded areas correspond to a ferromagnetic ground state. Dashed curves correspond (from top to bottom) to $\alpha_{J_{ex}}=0.2,0.4,0.6$, and  0.8. }
\label{2pla}
\end{center}
\end{figure}

The mobile atoms in the excited band are also  expected to stabilize, via double-exchange  processes,   the ferromagnetic phase
when a weak tunneling between plaquettes ($J'$ and $J_e'$) is allowed. The
stabilization mechanism relies
on the preservation of the spin when an atom hops, and on the energy penalty of $2J_{ex}$ when ground and excited atoms form a singlet instead of a triplet at a given site.  Only when the spins of adjacent plaquettes are fully aligned,  the mobile atoms are free to hop. We confirm the stabilization of the ferromagnetic correlations  by studying the  weakly coupled regime, which can be described again by an effective Heisenberg Hamiltonian as in Eq.~(\ref{hei}), but now between the $S=2$ states at each plaquette.

\begin{figure}[h]
\begin{center}
\begin{tabular}{cc}
\includegraphics[scale=0.55,angle=0]{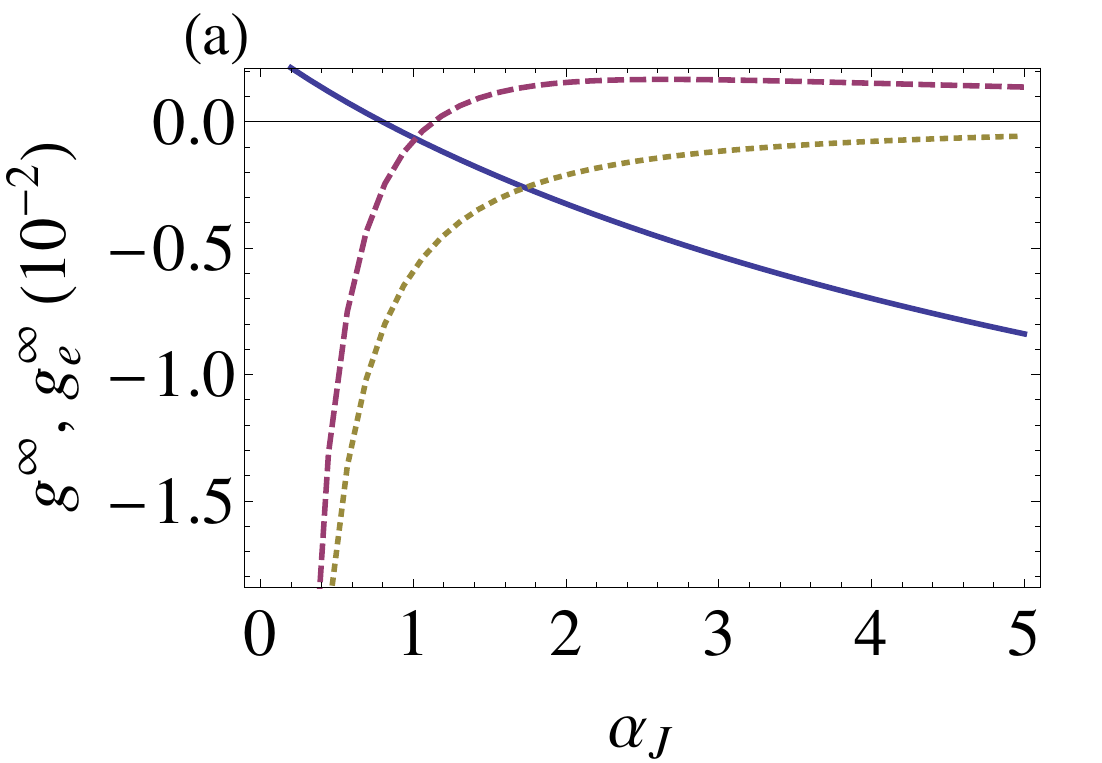}&
\includegraphics[scale=0.55,angle=0]{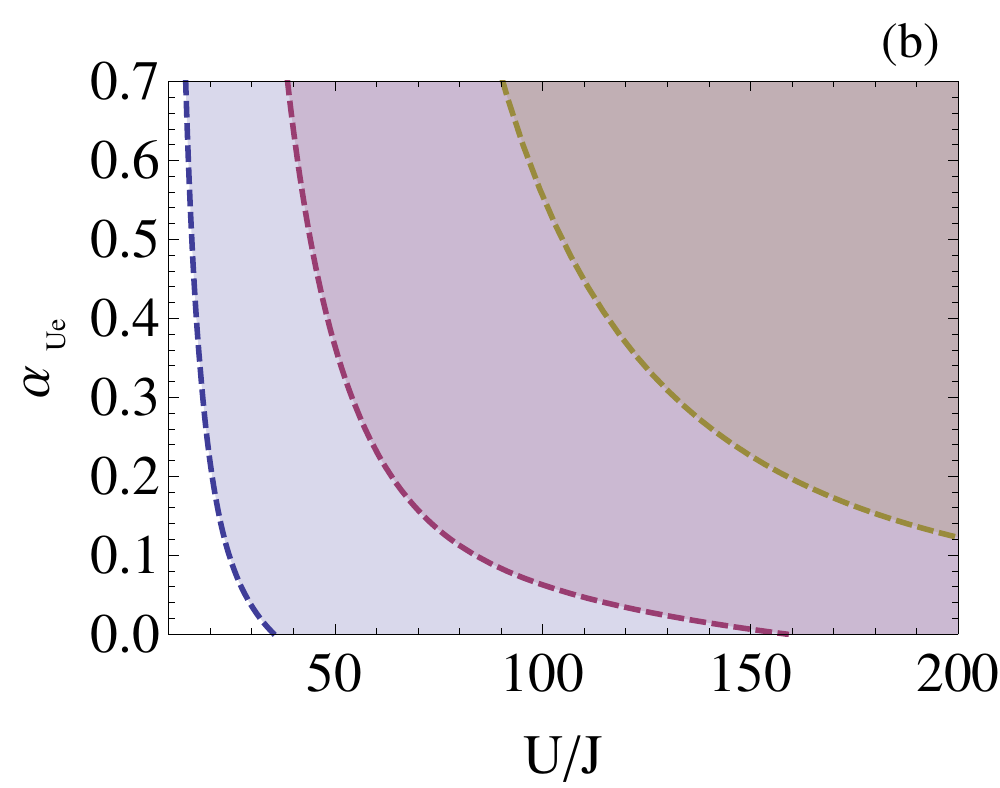}
\end{tabular}
\caption{ (a) Asymptotic coefficients $g^\infty$ and $g^\infty_e$ as functions of $\alpha_J$. The solid curve corresponds to $g$, and the dotted curve corresponds to $g_e$. The dashed curve corresponds to $g_e$ when $\alpha_{U_e}=0$. (b) Phase  diagram for $J'=0$. The shaded regions correspond to the ferromagnetic regime for different  $\alpha_{J}$ values. From left to right, $\alpha_{J}=0.5$, $\alpha_{J}=1.1$, $\alpha_{J}=2$.
}
\label{newstudies}
\end{center}
\end{figure}

 The coupling coefficient $G$  can be obtained by considering virtual processes in which one atom from the ground (excited) band hops from one plaquette to the other and returns to the original configuration. In practice, we extract $G$ from the analysis of the low-energy spectrum of two weakly coupled plaquettes [see  Fig.~\ref{2pla2band}].
 $G$ depends in a nontrivial way on the interaction and kinetic energy parameters and it is given by  $G=g J'^2/J+g_e J_e'^2/J_e$ for $J',J'_e\ll J, J_e$. Here  $g$ and $g_e$   are proportional to $G$ when the interplaquette tunneling is allowed only in the ground or excited band, respectively. Both   depend on $J$, $J_e$, $U$, and $J_{ex}$; and  $g_e$ also depends on $U_e$.
In the effective model, the existence of ferromagnetism can be directly deduced from the sign of $G$. To further understand the robustness of the ferromagnetic phase, we consider the general case for which all the interaction terms are proportional to each other, $U_e=\alpha_{U_e} U$, $V=J_{ex}=\alpha_{J_{ex}} U$, and the kinetic terms are related as $J_e=\alpha_J J$.
For large interaction values ($U\to\infty$), the coefficients $g$ and $g_e$ approach to their asymptotic values $g^\infty$ and $g^\infty_e$ that only depend on  $\alpha_J$ (independent of $\alpha_{J_{ex}}$ and $\alpha_{U_e}$ so long as $\alpha_{J_{ex}}>0$ and $\alpha_{U_e}>0$).  Figure~\ref{newstudies}~(a) shows that both $g^\infty$ and $g^\infty_e$  become negative for a large parameter regime, confirming the robustness of the ferromagnetic phase. If $U_e=0$ [dashed curve in Fig.~\ref{newstudies}~(a)], $g^\infty_e$ becomes positive for an important region of $\alpha_J$ values. This finding is consistent  with  Ref.~\cite{simon2007nsf} where it is pointed out that a nonzero interaction between atoms in the excited band ($U_e>0$) can be crucial for the transition to a ferromagnetic ground state.

Figure~\ref{newstudies}~(b)  shows  the phase diagram as a function of $\alpha_{U_e}$ and $U/J$  for three different $\alpha_J=0.5$, 1.1, 2 values, with the assumption that $J'=0$.  For this study, we set $\alpha_{J_{ex}}=0.4$, which is a typical value for optical superlattices. In general, we observe that as $\alpha_{J_{ex}}$ increases,  ferromagnetism becomes more favorable.
 The transition to a ferromagnetic ground state depends strongly on $\alpha_J$ and, for the $J'=0$ case, low values of $\alpha_J$ enhance ferromagnetism. %
For $J_e'=0$, the phase diagram is independent of $\alpha_{U_e}$, and higher mobility of the excited atoms, large  $\alpha_J$, favors ferromagnetic correlations. For example, for $\alpha_{J}=2$ the critical value $U_c$ is $\approx 147 J$. It increases to  $U_c\approx 488 J$ for $\alpha_{J}=1.1$, and  for $\alpha_{J}=0.5$, no ferromagnetic phase is observed.

 \begin{figure}[h]
\begin{center}
\includegraphics[scale=0.55,angle=0]{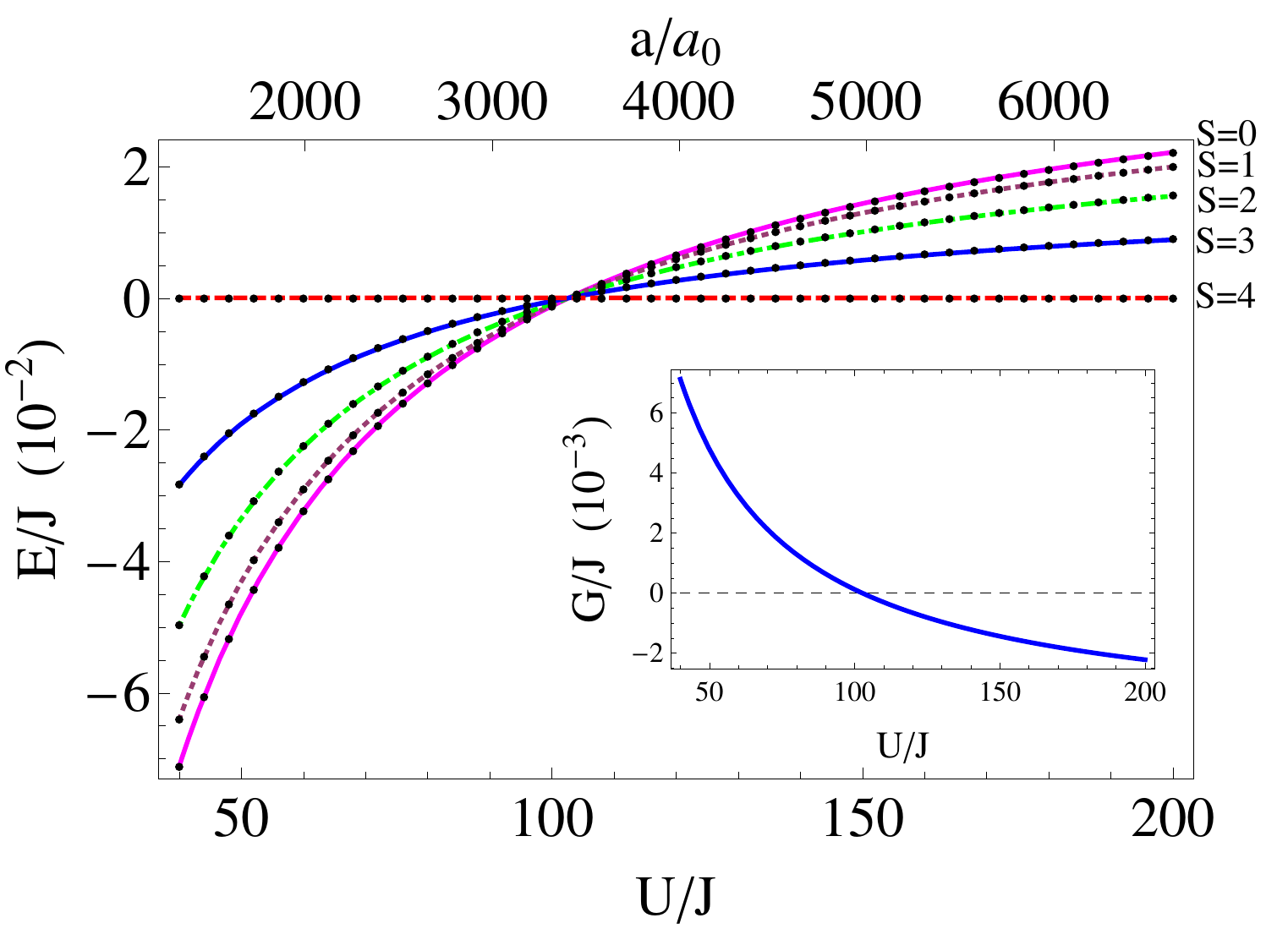}
\caption{
Lowest energies as a function of $U/J$ and the scattering length of two weakly coupled plaquettes ($V_l=4 E_r$ and  $V_s=5.5 E_r$) with six particles in the lowest band and two in the excited band and a total $S_z=0$. Circles correspond to exact numerical calculations, and lines correspond to the effective Hamiltonian [Eq.~(\ref{hei})] description. Top grid: scattering length values in Bohr radii for $^6$Li parameters.
Inset: The $G$ coefficient as a function of U/J for a plaquette with $V_l=4 E_r$ and $V_s=5.5 E_r$ (dashed curve).
For this case, the tunneling is $J\approx0.085 E_r$. The transition occurs at $U/J\approx100$. }
\label{res2p}
\end{center}
\end{figure}

 However, in realistic experimental setups it is not possible to explore the complete parameter space. For standard  superlattice geometries,
 in which only the ratio $V_l/V_s$ controls the relation between the different tunneling parameters,
we find that
a favorable ferromagnetic scenario takes place in the regime  when   $J=J_e$ and $J'=J'_e$.   We can achieve this scenario by
loading the atoms in the lowest vibrational state of a 2D plaquette array (tight confinement along $z$)
 creating a 2D lattice in the $x-y$ direction and
 and then exciting  one atom per plaquette to the first vibrational orbital along the $z$ direction. The atoms are initially all in the ground-state orbital $\Phi_g(\mathbf{r})=\phi_0(x)\phi_0(y)\widetilde{\phi}_0(z)$, and one of the atoms in each plaquette is excited to the $\Phi_e(\mathbf{r})=\phi_0(x)\phi_0(y)\widetilde{\phi}_1(z)$ vibrational state. The tunneling along the $x$ and $y$ directions is independent of the $z$-orbital ($\widetilde{\phi}_0(z)$ or $\widetilde{\phi}_1(z)$) and, therefore, $J=J_e$ and $J'=J'_e$. The tunneling in the $z$ direction depends on the $z$-orbital but it is negligible in the 2D geometry in consideration.

   Figure~\ref{res2p} presents this case for realistic  $^6$Li experimental parameters ($\alpha_{J_{ex}}=0.5$ and $\alpha_{U_e}=0.7$) where  we observe a change in the sign of the coupling coefficient $G$ (see inset) from positive to negative at $U/J\sim 100$, a value of interaction accessible with a Feshbach resonance. Even though for  the perturbative regime under consideration, where the effective Hamiltonian is valid,  $G$ is small ( of the order of few Hz), it is
   measurable with state-of-the-art  technology, as demonstrated  in  recent experiments where superexchange interactions as low as $5$ Hz have been resolved~\cite{Trotzky2008}.

For preparing and probing ferromagnetism in the isolated two-band plaquette array, we propose to start by loading atoms into a band insulator in a deep 3D lattice\cite{Jordens2008,Schneider2008}.
By applying a double-well superlattice along the z-direction and manipulating the double-well  bias, one can excite one of the four atoms in each double well to the first vibrational state via interaction blockade~\cite{cheinet2008cau}. Then, by slowly merging the double wells along the $z$-direction and splitting them along the $x-y$ directions, one can load the desired plaquette  configuration
with three atoms in the lowest and the fourth in the first-excited vibrational state along $z$.
 Alternatively, spatially selective two-photon Raman pulses
 can be applied to  excite one atom in each  plaquette  to the desired vibrational state~\cite{müller2007state,paredesPRA023603,gorshkov2008cqo}.
This procedure will lead to plaquettes with $S=0$. The state $S=0$ can be adiabatically converted into $S=1$ by applying a nonlinear  magnetic field gradient\footnote{ Our calculations show that a simple linear gradient does not couple the $S=1$ and $S=0$ states.} along $z$ and tuning the interactions from $U<0$ to $U>0$. The transition between $S=1$ and $S=2$ states} in a single two-band plaquette can be probed using the same experimental techniques proposed for the single band plaquette. A magnetic field gradient couples the $S=1$ and $S=2$ states [see inset of Fig.~\ref{2pla}(a)] and this coupling can be used to probe the transition with spin-imbalance measurements or band mapping techniques.

Interaction induced ferromagnetism in a few coupled plaquette array could also be observed by varying $U/J$ in the presence of a magnetic field gradient. Since our initially prepared state is a band  insulator with $S=0$ the  magnetic field gradient has to be large enough  to couple $S=0$ with $S=S_{max}$ across the transition.
After this procedure is applied the ferromagnetic nature of the ground state can be inferred in the applied  magnetic-field gradient by measuring  the local magnetization of the system~\cite{jo2009itinerant,weld2009spin,Babadi2009}. A linear magnetic-field gradient produces a perturbation in the effective Hamiltonian of the form $H_p=\sum_i i\delta E_p \hat{S}_{zi}/\hbar$, where $\delta E_p$ is the average energy shift between consecutive plaquettes. In  the ferromagnetic phase, the formation of a domain wall is expected.  The domain-wall width will be  determined by the dimensionless parameter $zGS/\delta E_p$, where $z$ is the number of nearest-neighbor plaquettes, and $S=2$. The measurement of this width can be used to extract $G$ in the ferromagnetic regime. In the antiferromagnetic phase on the contrary, no domain wall will be formed, and the local Neel order parameter should vary smoothly.
The onset of ferromagnetic correlations  as the system is driven
through the critical point should  be signaled by  a  suppression in
inelastic collisions, a minimum in kinetic
energy, and a maximum in the size of the cloud. The  latter
signatures have been demonstrated to be useful  smoking guns of a
ferromagnetic transition in recent experiments carried on in fermionic
gases without a lattice  (See Ref.~\cite{jo2009itinerant}  and references
therein).

\section{Conclusions}

 In summary, we have proposed a controllable and experimentally realizable scheme to study interaction-induced ferromagnetism in ultracold atoms.  Our method exploits the advantage offered by these systems  to divide the full lattice into  plaquettes.

We showed that a plaquette loaded with three fermionic atoms is a promising set-up to experimentally observe  Nagaoka ferromagnetism for the first time. We also used the plaquettes as the fundamental building blocks to create  interaction induced ferromagnetism at longer length scales.
We  analyzed the system of weakly coupled two-band plaquettes and demonstrated that in the limit where the interplaquette coupling can be treated perturbatively, the system maps out into an effective Heisenberg  Hamiltonian with a coupling constant $G$ which changes sign from positive to negative as interactions are increased. The change in sign is a manifestation of  an antiferromagnetic to ferromagnetic transition.

In the perturbative regime, where our analysis is valid, the coupling parameter $G$ is small. Consequently, it will be experimentally challenging to adiabatically reach the ferromagnetic ground state for large number of coupled plaquettes by increasing interactions. In this situation, the required temperature would be smaller than $G\sim 10^{-2}J$.
 Nevertheless, exact diagonalization in a two-plaquette array confirmed
 the persistence of itinerant ferromagnetism beyond the weakly coupling regime. Actually, we found  excellent agreement  between our perturbative Hamiltonian  and the many-body spectrum   even at values of $J'/J$ as high as $1/4$, as shown in the low energy spectrum presented in Fig.~\ref{res2p}.
 Our two-plaquette results seem to be consistent with variational Monte Carlo~\cite{kubo2009cms} and dynamical mean-field~\cite{held1998mcf} predictions that have found ferromagnetic phases  in  the two-band generic square-lattice  Hubbard model. The actual
  stability of the ferromagnetic phase in larger plaquette arrays and stronger interplaquette couplings  will need however to be resolved  ultimately by
experiments.


\ack
This work was supported by NSF, ITAMP, CUA and DARPA.


\section*{References}
\begin{thebibliography}{10}

\bibitem{PhysRev.142.350}
D.~R. Penn, Phys. Rev. {\bf 142},  350  (1966).

\bibitem{stoner1938cef}
E. Stoner, Proc. R. Soc. London {\bf 165},  372  (1938).

\bibitem{fazekas1990gsp}
P. Fazekas, B. Menge, and E. M{\"u}ller-Hartmann, Z. Phys. B {\bf 78},  69
  (1990).
A.~N. Tahvildar-Zadeh, J.~K. Freericks, and M. Jarrell, Phys. Rev. B {\bf 55},
  942  (1997).

\bibitem{Wu}
{Some examples of ferromagnetism in Hubbard models with complex lattice
  geometries have also been predicted: H. Tasaki, Phys. Rev. Lett. 75, 4678
  (1995), S. Zhang, H. Hung and C. Wu, arXiv:0805.3031},   .

\bibitem{Nagaoka1966}
Y. Nagaoka, Phys. Rev. {\bf 147},  392  (1966).

\bibitem{takahashi1982chm}
M. Takahashi, J. Phys. Soc. Japan {\bf 51},  3475  (1982).
Y. Fang {\it et al.}, Phys. Rev. B {\bf 40},  7406  (1989);
B. Doucot and X.~G. Wen, Phys. Rev. B {\bf 40},  2719  (1989).


\bibitem{Nielsen2007}
E. Nielsen and R.~N. Bhatt, Phys. Rev. B {\bf 76},  161202(R)  (2007).

\bibitem{Yao2007}
H. Yao, W.~F. Tsai, and S.~A. Kivelson, Phys. Rev. B {\bf 76},  161104  (2007).

\bibitem{hewson1993kph}
A. Hewson, {\em {The Kondo Problem to Heavy Fermions}} (Cambridge University
  Press, Cambridge, 1993).
H. Tsunetsugu, M. Sigrist, and K. Ueda, Rev. Mod. Phys {\bf 69},  809  (1997);
M. Gulacsi, Phil. Mag. {\bf 86},  1907  (2006).


\bibitem{Zener}
C. Zener, Phys. Rev. {\bf 81},  440  (1951).

\bibitem{Tasaki1998}
H. Tasaki, Progr. Theor. Phys. {\bf 99},  489  (1998).

\bibitem{Greiner2001}
M. Greiner, I. Bloch, O. Mandel, T.~W. Hansch, and T. Esslinger, Phys. Rev.
  Lett. {\bf 87}, 160405 (2001).

\bibitem{Trotzky2008}
S. Trotzky, P. Cheinet, S. Folling, M. Feld, U. Schnorrberger, A.~M. Rey, A.
  Polkovnikov, E.~A. Demler, M.~D. Lukin, and I. Bloch, Science {\bf 319},  295
   (2008).

\bibitem{simon2007nsf}
P. Simon and D. Loss, Phys. Rev. Lett. {\bf 98},  156401  (2007).

\bibitem{Jordens2008}
R. J\"ordens, N. Strohmaier, K. G\"unter, H. Moritz, T. Esslinger,  Nature (London) {\bf 455}, 204-207 (2008)


\bibitem{Schneider2008}
U. Schneider, L. Hackermuller, S. Will, T. Best, I. Bloch, T.~A. Costi, R.~W. Helmes, D. Rasch, A. Rosch, Science {\bf 322},  1520  (2008).


\bibitem{cheinet2008cau}
P. Cheinet, S. Trotzky, M. Feld, U. Schnorrberger, M. Moreno-Cardoner, S.
  Foelling, and I. Bloch, Phys. Rev. Lett. {\bf 101}, 090404 (2008).



\bibitem{müller2007state}
T. M{\"u}ller, S. F{\"o}lling, A. Widera, I. Bloch, Phys. Rev. Lett. {\bf 99}, 200405 (2007)

\bibitem{paredesPRA023603}
B. Paredes and I. Bloch, Phys. Rev. A {\bf 77},  023603  (2008).

\bibitem{gorshkov2008cqo}
A. Gorshkov, L. Jiang, M. Greiner, P. Zoller, and M. Lukin, Phys. Rev. Lett.
  {\bf 100},  093005  (2008).

\bibitem{jo2009itinerant}
G. B. Jo, Y. R. Lee, J. H. Choi, C. A. Christensen, T. H. Kim, J. H. Thywissen, D. E. Pritchard, W. Ketterle, Science {\bf 325}, 1521 (2009)


\bibitem{weld2009spin}
D. Weld, P. Medley, H. Miyake, D. Hucul, D. Pritchard, and W. Ketterle,
Phys. Rev. Lett. {\bf 103}, 245301 (2009)



\bibitem{Babadi2009}
M. Babadi, D. Pekker, R. Sensarma, A. Georges, E. Demler, Arxiv:0908.3483  (2009).


\bibitem{kubo2009cms}
K. Kubo, Phys. Rev. B {\bf 79},  020407  (2009).

\bibitem{held1998mcf}
K. Held and D. Vollhardt, Eur. Phys. J. B {\bf 5},  473  (1998).

\end{thebibliography}
\end{document}